\documentclass[a4paper,11pt]{article}
\usepackage{jheppub} 
\usepackage{lineno}

\usepackage{tikz}
\usetikzlibrary[snakes]
\usepackage{eufrak}
\usepackage{mathrsfs}
\usepackage{bbm}
\usepackage{mathabx}
\usepackage{tikz}
\usetikzlibrary[snakes]

\usepackage{graphicx}
\usepackage{subcaption}
\graphicspath{ {Images/} }

\title{\boldmath A free field approach to boundary $\widehat{g}_{k}$ WZW models.}

\author{Xun Liu}
\affiliation{Department of Physics, The University of Tokyo \\
7-3-1 Hongo, Bunkyo-ku, Tokyo 113-0033, Japan}

\emailAdd{xun.liu@tnp.phys.s.u-tokyo.ac.jp}

\abstract{The Wakimoto-type free-field approach is applied to the boundary integer-level simple $\widehat{g}(k)$ Wess-Zumino-Witten (WZW) models, with a renewed motivation. With the introduction of the Lauricella hypergeometric functions $F_{D}^{(n)}$ and their analytical extensions, we could obtain all genus-zero bulk $n$-point functions explicitly, for rational conformal field theories (RCFTs) that admit a free-field approach. I present free-field expressions for $\widehat{g}(k)$ Ishibashi states, and provide simple example calculations in the simplest models. An extreme short discussion on potential generalizations of free-field approach to the logarithmic WZW models at the admissible levels is also given.}

\begin{document}
\maketitle
\flushbottom

\section{Introduction}
\label{sec:intro}

The infinite-dimensional local conformal symmetries in two-dimensional field theories (CFT$_2$) lead to much deeper understandings of the topic, compared to generic quantum field theories \cite{Belavin:1984vu}. Free-field approaches have been efficient methods in the studies of both the correlation functions and the representation theories of the simplest rational CFT$_2$s (RCFTs) \cite{Dotsenko:1984nm, Dotsenko:1984ad, Dotsenko:1985hi, Felder:1988zp, Bernard:1989iy, Bouwknegt:1990wa}. Note that a CFT$_2$, not necessarily an RCFT, is considered to admit a free-field approach if the theory satisfies the following three conditions \cite{Bouwknegt:1990wa}
\begin{itemize}
    \item  The realization of its chiral algebra $\mathcal{A}$ and anti-chiral algebra $\widebar{\mathcal{A}}$ in terms of the free fields.

    \item  The existence of projection maps from the free-field Fock space modules to the chiral algebra representations in the CFT$_2$ Hilbert space, called the Fock space resolutions.

    \item  The existence of a method for calculating the correlation functions in the original CFT$_2$ using the free-field vertex operators, with insertions of the free-field $\mathcal{A}$ intertwiners. 
\end{itemize}

I revisit this topic with a renewed motivation. I proposed a method to challenge the calculation of genus-$g$, bulk $n$-point, $b$-boundary, $c$-crosscap functions with $x$-boundary operator $\mathcal{F}_{g,n,b,c}^{x}$ by expanding them as infinite linear combinations of genus-$g$, bulk $(n+b+c)$-point functions $\mathcal{F}_{g,(n+b+c)}$ \cite{Liu:2023gzf}. The expansion is realized by the insertion of $I \otimes \widebar{I}$, expressed by the complete basis of the CFT$_2$ Hilbert space, left to boundary or crosscap Ishibashi states
\begin{equation}
   \vert b,c \rangle =   \sum_{ \alpha , \beta }   \:  \vert \phi_{\alpha} \rangle  (G)^{\alpha \beta}  \langle  (\phi_{\beta})^{c}  \vert b,c \rangle .\label{eq; b c sta exp}
\end{equation}
The matrix $(G)^{\alpha \beta} $is the inversion Gram matrix. Obtaining the infinite set of linear coefficients in (\ref{eq; b c sta exp}) is complicated, especially for a free exceptions where the chiral algebras are simple, including the free-field theories.

I provide the procedure to obtain genus-zero bulk $n$-point functions explicitly, for all RCFTs that admit a free-field approach. The core of procedure is the introduction of the Lauricella hypergeometric function $F_{D}^{(n)}$ and their analytical extensions, where the analytical extensions are introduced to deal with endpoint divergency problem that might arise from the contour deformation to Euler-type integrals. The procedure for obtaining all RCFT genus-zero correlation functions is the repeated application of contour integrals and Talyor expansions of $F_{D}^{(n)}$, which is cumbersome, yet free of technical difficulties. I apply the Wakimoto-type free-field approach to the charge-conjugated and diagonal $\widehat{g}(k)$ Wess-Zumino-Witten (WZW) models, where $g$ being a simple Lie algebra \cite{Witten:1983ar, Knizhnik:1984nr, Gepner:1986wi}. I express the rational WZW Ishibashi states in terms of Fock space Ishibashi states, and calculate the disk correlation functions using generalized Coulomb-gas formalism in the simplest models. Despite of the early effort of applying the Coulomb-gas formalism to the calculations of the genus-zero correlation functions, new aspects are pointed out in this work

The work will be concluded with a short discussion on potential generalizations to the admissible level WZW models, which are irrational logarithmic CFT$_2$s, where many breakthrough discoveries has been made in the past decade \cite{Saleur:1991hk, Saleur:1991vh, Rozansky:1992td, Gurarie:1993xq, Creutzig:2013hma, Kawasetsu:2018lur, Kawasetsu:2019att}.

Some other conventional works that have discussed the applications of the free-field approaches to RCFTs are \cite{Parkhomenko:2001ki, Kawai:2002vd, Kawai:2002pz, Caldeira:2003zz, Parkhomenko:2003gy, Hemming:2004dm, Parkhomenko:2004ab}.

\

\section{Free-field approaches to boundary $\widehat{g}(k)$ Wess-Zumino-Witten (WZW) models}

\subsection{Preliminaries in Wess-Zumino-Witten (WZW) models}

Necessary concepts of simple Lie algebras $g$, their corresponding Kac-Moody algebras, integer level rational Wess-Zumino-Witten (WZW) models, and the Cardy boundary states in RCFTs are reviewed in this section.

\

\paragraph{Basic notations.}

The Kac-Moody (KM) algebras considered in this work are those related to the simple Lie algebras, which will be called simple KM algebras in this work. Systematic introductions of this topic are given in \cite{Kac:1984mq, Kac:1990gs}. A Kac-Moody algebra is a $\mathbb{Z}_{n}$ graded extension of finite-dimensional Lie algebra. Central extensions are allowed in the KM algebras. For the simple KM algebras, the central extension is unique, and it is called the level $k$. I denote simple KM algebras by $\widehat{g}(k)$.

A common basis for the simple KM algebras is the Cartan-Weyl basis. It consist of the Cartan generators $H^{i}_{n}$ $i=1, \cdots , r$, that are related to simple roots $\alpha_{i}$ of $g$, and the ladder generators $E^{\alpha}_{n}$ that correspond to the roots of $g$. The set of roots is denoted as $\Delta$, and $\Delta = \Delta_{+} \cup \Delta_{-}$, $\Delta_{-} = -\Delta_{+}$. The Cartan-Weyl basis of $g$ reveals the relation $\vert \Delta \vert + r = \dim (g)$. Simple Lie algebras $g$ have unique highest roots $\theta$ and it decompose as 
\begin{equation}
    \theta = \sum_{i=1}^{r} a_{i} \alpha_{i} = \sum_{i=1}^{r} a_{i}^{\vee} \alpha_{i}^{\vee} , 
\end{equation}
where $\alpha_{i}^{\vee} :=  2 \alpha_{i} / \vert \alpha_{i} \vert $ are simple coroots. The Coxeter number $h$ and the dual Coxeter number $h^{\vee}$ are defined as 
\begin{equation}
    h = 1+ \sum_{i=1}^{r} a_{i} \: , \quad   h^{\vee}= 1+ \sum_{i=1}^{r} a_{i}^{\vee} . 
\end{equation}
The Virasoro subalgebra of $\widehat{g}(k)$ constructed from the Sugawara construction 
\begin{equation}
    T^{g}(z) = \frac{1}{2(k+h^{\vee})}\Big\{ \sum_{i} : H^{i} H^{i}:(z) + \sum_{\alpha \in \Delta_{+} } \frac{\vert \alpha \vert^{2} }{2}  :  (E^{\alpha} E^{-\alpha})  : (z) \Big\}  , \label{eq: Sug war cons}
\end{equation}
whose Virasoro central charge being 
\begin{equation}
    c \big[ \widehat{g}(k)\big] = \frac{ k \dim (g) }{k+h^{\vee}} = r- \frac{12 \vert \rho \vert^{2} }{k+h^{\vee}} + \vert \Delta \vert  , \label{eq: gk Vir cen}
\end{equation}
where the Freudenthal-de Vries strange formula is applied in the second equal.

We choose the fundamental weights $\omega_{i}$ of $g$ as the basis of $g$ weight vectors $\lambda$
\begin{equation}
    \lambda = \sum_{i=1}^{r} \lambda_{i} \omega_{i} = [ \lambda_{1}, \cdots , \lambda_{r} ] , \quad  (\omega_{i}, \alpha^{\vee}_{j})= \delta_{ij} ,
\end{equation}
where the linear coefficients are called the Dynkin labels of $\lambda$. A $g$-weight is called dominant if $\lambda_{i} \in \mathbb{N}$, with at least one positive Dynkin label. The weight space of the simple KM algebra $\widehat{g}(k)$ is $(r+1)$-dimensional, with the zeroth Dynkin label being
\begin{equation}
    \lambda_{0} =  k- (\lambda, \theta)  = k - \sum_{i=1}^{r} a_{i}^{\vee} \lambda_{i} . 
\end{equation}
Similarly, a $\widehat{g}(k)$ weight is dominant if all its Dynkin labels are non-negative integers, with at least one being positive. It is obvious that dominant weights only exist for algebras with positive integer levels $k\in \mathbb{Z}_{+}$, and the set of level-$k$ dominant $\widehat{g}(k)$ weights is denoted by $\widehat{P}_{+}^{k}$.

Define a Weyl reflection of a $g$ weight $\lambda$ with respect to a root $\alpha$ as
\begin{equation}
    w_{\alpha}\lambda := \lambda - (\alpha^{\vee} ,  \lambda) \alpha .  
\end{equation}
The Weyl reflections of $g$ form a finite-dimensional group, called the Weyl group $W(g)$, generated by the simple Weyl reflections 
\begin{equation}
    w_{i} \lambda =  \lambda -( \alpha_{i}^{\vee} , \lambda ) \alpha_{i} = \lambda - \lambda_{i} \alpha_{i} . 
\end{equation}
The decomposition of a Weyl reflection $w\in W(g)$ in terms of simple Weyl reflections is not unique. We define the length $l(w)$ of $w$ as the minimal number of simple Weyl reflections of expressing $w$. There is a unique longest element $w_{l} \in W (g)$ and the charge conjugation of $\lambda$ is defined as $U \lambda : =- w_{l} \lambda $. The Weyl group of $\widehat{g}(k)$ are generated by $(r+1)$ simple Weyl reflections, and they are infinite-dimensional. It is proven that the Kac-Moody Weyl group $\widehat{W}(\widehat{g})$ is a semi-direct product of the finite Weyl group $W(g)$ and a translation group $T$ along the coroot lattice $Q^{\vee}$ of $g$, generated by translations $k \alpha_{i}^{\vee} $
\begin{equation}
    \widehat{W}(\widehat{g})= W (g) \ltimes T .
\end{equation}
The shifted Weyl transformations are also defined analogously to the finite cases.

The irreducible modules in the Hilbert spaces of the integral-level $\widehat{g}(k)$ dominant modules, with the elimination of singular submodules. The embedding structure of the dominant modules are given by the Kac-Weyl formula, and it is a standard form where the singular vectors taking the form of $\widehat{w}\ast \widehat{\Lambda}$. The integer level $\widehat{g}(k)$ WZW models are RCFTs, and hence their Hilbert spaces are constrained by the torus partition function modular invariant condition \cite{Cardy:1986ie}. Obtaining all modular invariants and classifying is a highly challenging question. In this work, we only focus on the models with the diagonal and charge-conjugated modular invariants, which exist for all RCFTs \cite{Moore:1988qv}.

\ 

\paragraph{The Wakimoto free-field approach.}

The form of the KM Virasoro central charge $c ( \widehat{g}(k) \big]$ indicated that the same Virasoro central charge can be realized by $r$ background charged scalar bosons $\phi^{i}$ and $\vert \Delta_{+} \vert$ $\beta^{\alpha} \gamma^{\alpha}$, $\alpha\in \Delta_{+}$, bosonic ghost fields \cite{Wakimoto:1986gf, Bernard:1989iy, Bouwknegt:1989xa, Bouwknegt:1989jf, Bouwknegt:1990wa}
\begin{equation}
    T^{g}(z) =  -\frac{1}{2 }  : \partial \phi \cdot \partial \phi  : (z) - \alpha_{+} \rho \cdot i \partial^{2} \phi (z) - \sum_{\alpha \in \Delta_{+} } :  \beta^{\alpha} \partial  \gamma^{\alpha} : (z) ,
\end{equation}
where $\alpha_{+}^{2} = 1/ (k+h^{\vee}) $. The Cartan generators in the Chevalley basis are given by 
\begin{equation}
    h^{i}(z) = \sum_{\alpha \in  \Delta_{+}} ( \alpha, \alpha_{i}^{\vee})  : \gamma^{\alpha} \beta^{\alpha}  : (z)  + \alpha_{+}^{-1} \alpha_{i}^{\vee} \cdot i \partial \phi^{i} .
\end{equation}
The screening operators take the form of \cite{Bouwknegt:1990wa}
\begin{equation}
    \widetilde{s}_{i}^{+} (z)  = \rho (e_{i}) \: : e^{-i \alpha_{+} \alpha_{i} \cdot \phi } : (z) ,
\end{equation}
where $\rho (e_{i})$ is some polynomials of the $\beta^{\alpha}\gamma^{\alpha}$ fields \cite{Bouwknegt:1990wa}. Their explicit forms will be given when necessary. The KM intertwiners, which are closed contour integrals over the screening operators, are denoted by $\widetilde{d}$.

The Fock space modules in the Wakimoto-type have the triangular decomposition of the free-field algebra is $n_{+} = \{ \beta_{n \ge 0} , \: \gamma_{n>0} , \: a_{n>0}^{i} \}$, $n_{-} = \{ \beta_{n < 0} , \: \gamma_{n\le 0} , \: a_{n<0}^{i} \}$, and $h= \{ a^{i}_{0} \}$, with $\forall i , \: \forall \alpha\in \Delta_{+}$. The $\phi^{i}\beta^{\alpha} \gamma^{\alpha}$ Fock space modules $\widetilde{F}_{\widehat{\Lambda}}$ are labeled by a $r$-dimensional momentum $p$ (related to Dynkin labels by $\alpha_{i}^{\vee} \cdot p = \alpha_{+} \Lambda_{i} $). The form of resolution of dominant $\widehat{g}(k)$ modules is hinted from the form of the Kac-Weyl character formula, which can be written as  \cite{Bouwknegt:1989xa}
\begin{equation}
      \chi_{ \Lambda } = \sum_{ \widehat{w}  \in \widehat{W} } (-1)^{l(\widehat{w})} \text{Tr}_{ F_{ \widehat{w} \ast\Lambda } }  \big[ q^{L_{0}} e^{2\pi i  (z \cdot  H) }  \big] . \label{eq: Kac Weyl Fock exp}
\end{equation}
Hence, we consider a complex of the $\phi^{i}\beta^{\alpha} \gamma^{\alpha}$ Fock space modules \cite{Bernard:1989iy}
\begin{equation}
    \widetilde{C}_{\widehat{\Lambda}}: \quad  \cdots \overset{\widetilde{d}}{\longrightarrow} \widetilde{F}_{ \widehat{\Lambda} }^{(-1)}\overset{\widetilde{d}}{\longrightarrow} \widetilde{F}_{ \widehat{\Lambda} }^{(0)} \overset{\widetilde{d}}{\longrightarrow} \widetilde{F}_{ \widehat{\Lambda} }^{(1)} \overset{\widetilde{d}}{\longrightarrow} \cdots \:, 
\end{equation}
where $\widetilde{F}_{ \widehat{\Lambda} }^{(i)}$ is the direct sum over Fock space modules generated by highest weights obtained from shifted KM Weyl actions of elements with fixed length $i$
\begin{equation}
    \widetilde{F}_{ \widehat{\Lambda} }^{(i)} = \bigoplus_{ \widehat{w}  }^{l(\widehat{w}) =i} \: \widetilde{F}_{ \widehat{w} \ast \widehat{\Lambda} } .
\end{equation}
The differentials in the complex are combinations of $\widehat{g}(k)$ intertwiners that shift the primary $\widehat{g}(k)$ weights by the desired amount, ensuring that their are mapped to singular vectors in the dominant $\widehat{g}(k)$ Verma modules. The resolution conjecture states that the zeroth cohomology space of the complex is isomorphic to the dominant module $L_{ \widehat{\Lambda}}$, while all other cohomology spaces being trivial 
\begin{equation}
    H^{i} ( \widetilde{C}_{\widehat{\Lambda}} ) \quad \cong \quad  \begin{cases} \varnothing , \quad  i \ne 0 ; 
        \\  L_{\widehat{\Lambda}} , \quad  i=0 .
    \end{cases}  
\end{equation}
The vertex operator $V_{\widehat{\Lambda}}$ corresponding to the highest weight operators of the dominant modules are
\begin{equation}
    : e^{ i \alpha_{+} \Lambda \cdot \phi^{i}  } :  (z) ,
\end{equation}
which is easy to verify that it produce the correct primary eigenvalues (conformal weight $h$ and Dynkin labels) under the action of the free-field generators. Note that acting operator $\gamma$ will not change the conformal weights, hence the primary operators are obtained in such manner. In this work, our focus only on the highest-weight operators.

To obtain the Cardy boundary states, it is necessary to know the explicit forms of the modular $S$ matrices of the theory. The results are \cite{Kac:1984mq}
\begin{equation}
    S_{\widehat{\Lambda} \widehat{\Lambda}' } = \frac{i^{ \vert \Delta_{+} \vert }  \vert P/Q^{\vee} \vert^{-\frac{1}{2}}  }{\sqrt{(k+g)^{r}}} \sum_{w \in W(g)} \epsilon(w) e^{ - \frac{2\pi i}{k+g} ( w  (\Lambda+\rho) , \Lambda'+\rho ) } , \label{eq: domi S}
\end{equation}
where $P$ and $Q^{\vee}$ are the weight and coroot lattice of $g$ respectively.

The fusion rules of the WZW models are often calculated using the Kac-Walton formula \cite{Walton:1989sc}. However, in our work, we need to calculate the modular $S$ matrices in each model to obtain the form of the Cardy boundary states, hence, we apply the Verlinde formula to obtain the fusion rules \cite{Dijkgraaf:1988tf, Verlinde:1988sn}
\begin{equation}
    N_{\widehat{\Lambda}\widehat{\Lambda}'}^{\widehat{\Lambda}''} = \sum_{\widehat{\mu}} \frac{ S_{ \widehat{\Lambda} \widehat{\mu} } S_{ \widehat{\Lambda}' \widehat{\mu} } S_{ \widehat{\Lambda}'' \widehat{\mu} }^{-1} }{S_{ 0 \widehat{\mu} }}  .
\end{equation}

\

\subsection{Preliminaries in boundary RCFTs}

The presence of boundaries reduce the $\mathcal{A} \otimes \mathcal{A}$ bulk conformal symmetry to an open-sector chiral symmetry $\mathcal{A}' \subset \mathcal{A}$ \cite{Cardy:1984bb, Recknagel:1997sb, Recknagel:2013uja}. If $\mathcal{A}' = \mathcal{A}$, the boundary condition is called chiral symmetry preserved. Otherwise, they are called chiral symmetry violated. In this work, only chiral symmetry preserved boundary conditions are considered.

Denote the chiral generators by $\{ T(z) , W^{i}(z) ,\cdots \}$, they satisfy the gluing conditions on the upper half plane (UHP) \cite{Cardy:1984bb}
\begin{equation}
    T(z) = \widebar{T}(\widebar{z}) , \quad W^{i}(z) = \Omega \widebar{W}^{i}(\widebar{z}) ,
\end{equation}
where $\Omega$ is an outer-automorphism of $\mathcal{A}$. We map the UHP to a unit disk and assign a boundary state to its boundary. The gluing conditions on the UHP are transformed into 
\begin{equation}
    (L_{n}- \widebar{L}_{-n}) \vert b \rangle_{\Omega} = \big[ W_{n}^{i}- \Omega (-1)^{h_{i}} \widebar{W}^{i}_{-n}) \big] \vert b \rangle_{\Omega}  =0 , \quad \forall n \in \mathbb{Z},\: i  . \label{eq: Ishi con b}
\end{equation}
The solutions of (\ref{eq: Ishi con b}) is spanned by the $\Omega$-twisted boundary Ishibashi states $\vert  i ,b \rangle \rangle_{\Omega} $ \cite{Ishibashi:1988kg, Onogi:1988qk}
\begin{equation}
    \vert i ,b \rangle \rangle_{\Omega} = \sum_{N=0}^{\infty} \vert i,N \rangle \otimes  V_{\Omega} U \vert i,N \rangle   ,
\end{equation}
where $\vert i,N \rangle$ denotes the orthonormal basis of the irreducible module $L_{i}$ of $\mathcal{A}$. $V_{\Omega}$ is a unitary isomorphism induced from $\Omega$ such that $V_{\Omega} : \: L_{i}  \to L_{\Omega (i)} $, and it is a fusion rule automorphism $N_{ij}^{k}=N_{\Omega(i)\Omega(j)}^{\Omega(k)}$ \cite{Recknagel:2013uja}. The overlaps between the $\Omega$-twisted Ishibashi states are given by
\begin{equation}
    _{\Omega}\langle \langle i,b  \vert q^{L_{0} + \widebar{L}_{0} - \frac{c_{Vir}}{12}} \vert j,b \rangle \rangle_{\Omega}  = \delta_{ij} \chi_{i}(q^{2}) .
\end{equation}
The Ishibashi states are not physically admissible boundary states in RCFTs.

The logic to obtain physical boundary states is to consider calculating annulus partition functions by two distinct yet equivalent methods, equating them, and the form of the open-sector Hilbert spaces will provide strong constraint conditions to the form of boundary states \cite{Cardy:1989ir}. The details are explained in a wide range of works and it is omitted here, we consider the case of Cardy boundary states \cite{Cardy:1989ir}, where the boundary conditions have the same labels as chiral irreducible modules, and their forms are 
\begin{equation}
     \vert b_{\widehat{\Lambda}'} \rangle_{\Omega} = \sum_{ \widehat{\Lambda} } \frac{S_{\widehat{\Lambda}'\widehat{\Lambda}}}{\sqrt{S_{0\widehat{\Lambda}}}} \vert   L_{\widehat{\Lambda}} \rangle \rangle_{\Omega} .
\end{equation}

\ 

\section{The free-field Ishibashi states}

The free-field realizations of the $\widehat{g}(k)$ KM currents in terms of the Wakimoto-type $\phi^{i} \beta^{\alpha} \gamma^{\alpha}$ fields are complicated. However, the $\widehat{g}(k)$ trivial gluing conditions transform into the trivial $\phi^{i} \beta^{\alpha} \gamma^{\alpha}$ gluing conditions. Hence, in the charge conjugated $\widehat{g}(k)$ WZW models, the free-field Ishibashi states can be written as 
\begin{equation}
    \vert L_{\widehat{\Lambda}} \rangle\rangle_{I} = \sum_{\widehat{w} \in \widehat{W}} \theta_{\widehat{w}} \vert F_{ \widehat{w} \ast \widehat{\Lambda}}   \rangle \rangle_{I}\: \otimes \: \vert F^{\beta \gamma} \rangle \rangle_{I}
\end{equation}
where
\begin{equation}
    \vert F_{\widehat{\Lambda}}   \rangle \rangle_{I} = \exp{ \Big[ \sum_{n \in \mathbb{Z}_{+} } -\frac{1}{n} a_{-n} \cdot \widebar{a}_{-n}  \Big] } \vert p \otimes \widebar{p} \rangle , \quad \widebar{p} = -p - 2\alpha_{+} \rho ,
\end{equation}
\begin{equation}
     \vert F^{\beta \gamma} \rangle \rangle_{I} =  \exp{\Big[ \sum_{\alpha} \sum_{n \in\mathbb{Z} }  ( \beta^{\alpha}_{-n}  \widebar{\gamma}_{-n} -\gamma_{-n} \widebar{\beta}_{-n}  )\Big]} \vert 0 \otimes \widebar{0} \rangle ,
\end{equation}
and $\theta_{\widehat{w}}$ is an undetermine phase factor assigned to each Weyl group element $\widehat{w}$ to correctly reproduce the chiral characters when calculating the overlaps. The anti-holomorphic momentum $\widebar{p}=-p-2\alpha_{+}\rho$, which is the result of imposing the Virasoro gluing conditions.

For the diagonal models, the conjugated gluing automorphisms transforms into conjugated gluing conditions for $\phi^{i}$, with the $\beta^{\alpha} \beta^{\alpha}$ part remains unchanged. Hence, we write
\begin{equation}
    \vert L_{\widehat{\Lambda}} \rangle\rangle_{U} = \sum_{\widehat{w} \in \widehat{W}} \theta_{\widehat{w}} \vert F_{ \widehat{w} \ast \widehat{\Lambda}}   \rangle \rangle_{U}\: \otimes \: \vert F^{\beta \gamma} \rangle \rangle_{I} , 
\end{equation}
where the $\phi^{i}$ Ishibashi state is now  
\begin{equation}
    \vert F_{\widehat{\Lambda}}   \rangle \rangle_{U} = \exp{ \Big[ \sum_{n \in \mathbb{Z}_{+} } \frac{1}{n} a_{-n} \cdot  w_{l} \cdot \widebar{a}_{-n}  \Big] } \vert p \otimes \widebar{p} \rangle , \quad \widebar{p} = w_{l} \cdot (  p + \alpha_{+} \rho ) - \alpha_{+} \rho .
\end{equation}
By replacing the WZW Ishibashi states in (\ref{eq; b c sta exp}) with the free-field Ishibashi states, allows us to obtain the expansion coefficients, using the form of inverse Gram matrices and Ishibashi states of the $\phi^{i}\beta^{\alpha} \gamma^{\alpha}$ free-field theory.

\

\section{Correlation function computations}

A general description of how to use the generalized Coulomb-gas formalism to calculate RCFT correlation functions is provided. Then, some simple example of disk correlation function calculations are provided, focusing on revealing the crucial importance of the Lauricella hypergeometric function $F_{D}^{(n)}$.

\subsection{Generic description}

I begin with a generic description of the Coulomb-gas formalism \cite{Dotsenko:1984nm, Dotsenko:1984ad, Dotsenko:1985hi}. Correlation functions in CFT$_2$ can be obtained by solving the differential equations induced from inserting singular operators into correlation functions \cite{Knizhnik:1984nr, Belavin:1984vu}. This method works especially well for genus-zero surfaces, and when the chiral algebras are relatively simple and the number of the chiral operators are relatively low. With the growth of the number of the operators or the complexities of the chiral algebras, obtaining and solving the differential equations will become significantly more difficult, such that a more efficient method is desired.

The Coulomb-gas formalism was first proposed in the works \cite{Dotsenko:1984nm, Dotsenko:1984ad, Dotsenko:1985hi}, where the chiral correlation functions are calculated by using the free-field vertex operators, with insertions of Virasoro intertwiners, to satisfy the neutrality conditions of the vertex operators. A brief understanding of the Coulomb-gas formalism is the following: we use the vertex operators to represent chiral operators in the original theory. However, the fusion rules of the original theory differ from those of the free-field theory. Hence, we need a method to match the fusion rules on both sides. Inserting free-field $\mathcal{A}$ intertwiners is the perfect fit of our objectives, since the $\mathcal{A}$ representation theoretic aspects remain unchanged under the actions of $\mathcal{A}$ intertwiners, yet the screening operators are capable of shifting the original vanishing vertex correlation functions into non-vanishing ones.

In the calculations of the disk correlation functions, both holomorphic and anti-holomorphic sectors, we place screening operators and vertex operators in the following ordering 
\begin{equation}
   0= \vert  z_{n} \vert < \vert z^{(n)} \vert  < \vert z^{(n-1)} \vert <  \cdots  < \vert z^{(1)} \vert  < \vert z_{n-1} \vert < \vert z_{n-2} \vert < \cdots  <1 ,
\end{equation}
with a reversed ordering in the anti-holomorphic sector. The orders of the integrals are $dz^{(n)} \to dz^{(n-1)} \to \cdots \to dz^{(1)}$, and $d\widebar{z}^{(n)} \to d\widebar{z}^{(n-1)} \to \cdots \to d\widebar{z}^{(1)}$ respectively. The disk integrands take the form of 
\begin{equation}
    \delta \widebar{\delta} \prod_{i<j} z_{ij}^{\alpha_{ij}}\prod_{\widebar{i}<\widebar{j}} \widebar{z}_{\widebar{i}\widebar{j}}^{ ^{\alpha_{\widebar{i}\widebar{j}}}  } \prod_{i , \widebar{j}} (1- z_{i} \widebar{z}_{\widebar{j}})^{\alpha_{i \widebar{j} }^{\Omega}} , 
\end{equation}
where $\delta$ and $\widebar{\delta}$ denote the neutrality conditions in the holomorphic and anti-holomorphic sectors respectively. When performing the first $z^{(n)}$ integral, $z^{(n)}$ becomes the variable, the adjacency two branching point are $z_{2}$ and $z^{(n-1)}$, and we deform the contour along the branching cut, by going through the branching cut twice along different directions, and we obtain
\begin{equation}
  \int_{0}^{z^{(n-1)}} dz^{(n)}\:\cdots =  (1-e^{2\pi a})(1-e^{2\pi i (c-a) }) \oint  dz^{(n)}\:\cdots . 
\end{equation}
We omit the overall factors $ (1-e^{2\pi a})(1-e^{2\pi i (c-a) })$ in all our results in this work. After the performing the $z^{(n)}$ integral, we following the same procedure to deform the contour integral onto the segment $[0,z^{(n-2)}]$, and for all the rest of the integrals. Hence, the contour deformation we perform is 
\begin{equation}
    \oint dz^{(1)} \oint dz^{(2)}  \cdots \oint dz^{(n)} \quad \longrightarrow \quad \int_{0}^{z_{1}} dz^{(1)}   \int_{0}^{z^{(1)}} dz^{(2)} \cdots \int_{0}^{z^{(n-1)}} dz^{(n)} ,
\end{equation}
with all the factors are omitted. Similar procedure applies to the anti-holomorphic sector. The function at its core is the Lauricella hypergeometric functions, which include a few different classes of hypergeometric functions introduced as multi-variable extension of the Gaussian hypergeometric functions $_{1}F_{2}$. The Lauricella hypergeometric function $F_{D}^{(n)}$ admit Taylor expansions 
\begin{equation}
    F_{D}^{(n)} \Big[ a , b_{1} ,\cdots , b_{n} ; c ; z_{1} ,\cdots z_{n} \Big] = \sum_{k \in \mathbb{N} } \frac{  (a)_{ \sum_{i} k_{i} } \prod_{i=1}^{n} (b_{i})_{k_{i}}  }{ (c)_{ \sum_{i} k_{i} } }  \prod_{i=1}^{n} \Big( \frac{z_{i}}{k_{i}!} \Big)^{k_{i}} ,
\end{equation} 
with convergent radii being $ \vert z_{1} \vert<1$, $\forall i$. This function also has a Euler-type integral expression
\begin{equation}
     F_{D}^{(n)} \Big[ a , b_{1} ,\cdots , b_{n} ; c ; z_{1} ,\cdots z_{n} \Big] =  \frac{\Gamma(c)}{\Gamma(a)\Gamma(c-a)} \int_{0}^{1} dt \: (t)^{a-1} (1-t)^{c-a-1} \prod_{i=1}^{n} (1-t z_{i})^{-b_{i}},
\end{equation}
with $\text{Re}(a)>0$ and $\text{Re}(c-a)>0$. If the variables don't satisfy the conditions above, the integral has endpoint divergence problem.

Unfortunately, for the rational integer level $\widehat{g}(k)$ WZW models, the endpoint divergence is common, making the Coulomb-gas formalism seemingly inaccessible to most correlation functions. However, there should not be a divergent problem. For those with endpoint divergent problems, I avoid the contour deformation, and dealing with the contour integral by defining the extended Lauricella hypergeometric functions as 
\begin{equation}
    F_{D}^{(n)} \Big[ a , b_{1} , \cdots , b_{n}  ; c ; z_{1} ,\cdots , z_{n}  \Big]  :=  \frac{\Gamma(c)}{\Gamma(a)\Gamma(c-a)} \oint dt \:  t^{a-1} (1-t)^{c-a-1} (1-t z_{i})^{-b_{i}} ,
\end{equation}
whose Taylor expansions are 
\begin{equation}
    F_{D}^{(n)} \Big[ a , b_{1} ,\cdots , b_{n} ; c ; z_{1} ,\cdots z_{n} \Big] = \sum_{k \in \mathbb{N} }  c(a,b_{i};c;k_{i})  \prod_{i=1}^{n} (z_{i} )^{k_{i}} ,
\end{equation} 
with $ c(a,b_{i};c;k_{i})$ being unknown coefficients.

From the above setting, we claim that we obtained a systematic method to calculate all genus-zero bulk correlation functions, for RCFTs that admits a free-field approach, and obtaining the results is a repeated integration and Talyor expansion process. Comparing to the technically difficult solving the differential equation approach, this is a systematic approach without any technical difficulties, with the only disadvantages being that the cumbersomeness in most integrals, and the coefficients $ c(a,b_{i};c;k_{i})$ remains unknown.

\

Note that, that solutions to the neutrality conditions are usually not unique, especially for RCFTs like $\mathcal{W}$ minimal models, which contains Weyl symmetric identifications in their Kac-tables \cite{Bouwknegt:1992wg}. I claim that this is not an inconsistency in the Coulomb-gas formalism method, but different solutions to the neutrality conditions lead to different solutions to the differential equations induced from the insertion of singular operators \cite{Liu:2025dog}.

\

Before heading towards the detailed calculations of the disk correlation functions, we need to adjust one aspect. That is, when a single intertwiner is inserted, the chiral one-point function of the polynomial $\rho(z)$ will provide singular contribution to the integrand, not a complex constant. First, cases with only one intertwiner insertion happens only in the $\widehat{A}_{1}$ theories, since when the rank is higher than $1$, the simple roots of simple Lie algebras always contain negative Dynkin labels, and hence the neutrality condition is not possible to be satisfied with a single intertwiner insertion. For the $\widehat{A}_{1}(k)$ theories, $\rho= \beta(z)$. Then, the disk chiral one-point function provides an non constant contribution to the integrands. To obtain it, we consider
\begin{equation}
    \langle 0 \vert \beta(z) \exp{ \Big[ \sum_{n \in \mathbb{N}}  (\beta_{-n} \widebar{\gamma}_{-n} + \gamma_{-n} \widebar{\beta}_{-n} ) \Big]}  \vert 0  \rangle ,
\end{equation}
by expanding the exponential term and we obtain that the leading term provides a $z^{-1}$ term, with other terms being regular at the $z\to 0$ limit.

\

\subsection{Disk one-point functions}

The disk one-point functions are calculated without any intertwiner insertions. We only write down the integral results of the contributions from each Ishibashi states, without writing down the explicit form of the final summations of the disk correlation functions. However, it still provide the correct cross-boundary term in the straight-forward manner. The disk one-point functions for the charge-conjugated $\widehat{g}(k)$ models are 
\begin{equation}
   \langle b_{\widehat{\Lambda}'} \vert \Phi_{\widehat{\Lambda}}(z_{1},\widebar{z}_{1})  \vert 0 \rangle =  \frac{S_{\widehat{\Lambda}'\widehat{\Lambda}}^{-1}}{\sqrt{S_{0\widehat{\Lambda}}}} (1- \vert z_{1} \vert^{2} )^{ - \alpha_{+}^{2} \Lambda \cdot ( \Lambda+2\rho )  } = \frac{S_{\widehat{\Lambda}'\widehat{\Lambda}}^{-1}}{\sqrt{S_{0\widehat{\Lambda}}}} (1- \vert z_{1} \vert^{2} )^{ - 2h_{\widehat{\Lambda}}} ,
\end{equation}
which is proportional to the sphere chiral two-point functions, as expected. The disk one-point functions for the diagonal $\widehat{g}(k)$ models are 
\begin{equation}
   \langle b_{\widehat{\Lambda}'} \vert \Phi_{\widehat{\Lambda}} (z_{1},\widebar{z}_{1})  \vert 0 \rangle =  \frac{S_{\widehat{\Lambda}'\widehat{\Lambda}}^{-1}}{\sqrt{S_{0\widehat{\Lambda}}}} (1- \vert z_{1} \vert^{2} )^{  \alpha_{+}^{2} \Lambda \cdot  w_{l} \cdot (- w_{l} \cdot \Lambda+2\rho )  }.
\end{equation}

\

\subsection{$\widehat{A}_{1}(k)$ disk two-point and three-point functions}

Now we are ready to consider calculations where insertions of the intertwiners are required. The $\widehat{A}_{1}(k)$ modular $S$ matrix are
\begin{equation}
    S_{ \Lambda \mu } = \sqrt{\frac{2}{k+2}} \sin{\Big[ \frac{\pi (\Lambda_{1}+1) (\mu_{1}+1) }{(k+2)} \Big]}.
\end{equation}
The complete $\widehat{A}_{1}(k)$ fusion rules are given by unified by the Kac-Walton formula 
\begin{equation}
    N_{\Lambda \mu}^{\nu} = \begin{cases}
        1 , \quad  \vert \Lambda_{1}  - \mu_{1} \vert \le \nu_{1} \le \min\{ \Lambda_{1} + \mu_{1} , 2k-\Lambda_{1} - \mu_{1}  \},\: \text{and} \:  (\Lambda_{1} + \mu_{1}+ \nu_{1}) \in 2\mathbb{Z}; \\  0 , \: \text{otherwise} . 
    \end{cases}
\end{equation}

\ 

\paragraph{Level-$1$.}

The Virasoro central charge of the theory is $c= 1$. The level-$1$ Kac-table is $\widehat{\Lambda}=[1,0]$ and $[0,1]$. The modular $S$ matrix of this model are
\begin{equation}
    S=\frac{1}{\sqrt{2}} \begin{pmatrix}
        1 & 1 \\ 1 & -1
    \end{pmatrix} , 
\end{equation}
satisfying $S^{2}=C=I$. The fusion rule is 
\begin{equation}
    L_{[0,1]} \times L_{[0,1]} = L_{[1,0]} .
\end{equation}

We consider two-point functions $\langle b_{\widehat{\Lambda}} \vert \Phi_{[0,1]} (z_{1} ,\widebar{z}_{1})   \Phi_{[0,1]} (z_{2} ,\widebar{z}_{2}) \vert 0 \rangle $. A free-field realization of the contribution $I_{[0,1][0,1]}^{[1,0]}$ is 
\begin{equation}
    \oint d z \:   \widebar{V}_{[0,1]}^{\dag} (\widebar{z}_{2}) \widebar{V}_{[0,1]} (\widebar{z}_{1})    \widetilde{s}_{1}^{+} ( z)  V_{[0,1]} (z_{1}) V_{[0,1]} (z_{2})  ,
\end{equation}
The result includes the analytically extended Gaussian hypergeometric function
\begin{equation}
    \vert z_{1} \vert^{-1} (1- \vert z_{1} \vert^{2})^{\frac{1}{6}} \frac{\Gamma(-\frac{1}{3})\Gamma(\frac{2}{3})}{\Gamma(\frac{1}{3})} \:  _{1}F_{2} \Big[ -\frac{1}{3} , \frac{1}{3} ; \frac{1}{3} ;  \vert z_{1} \vert^{2} \Big] .
\end{equation}

\

We use the simplest $\widehat{A}_{1}(1)$ model to demonstrate the power of free-field approaches when the number of the bulk operators increase $\langle b_{\widehat{\Lambda}} \vert \Phi_{[0,1]} (z_{1} ,\widebar{z}_{1})   \Phi_{[0,1]} (z_{2} ,\widebar{z}_{2}) \Phi_{[0,1]} (z_{3} ,\widebar{z}_{3})\vert 0 \rangle$, containing the contribution $I_{[0,1][0,1][0,1]}^{[0,1]}$. A free-field realization of the $I_{[0,1][0,1][0,1]}^{[0,1]}$ is 
\begin{equation}
   \oint dz  \oint d\widebar{z} \: \widebar{V}_{[0,1]}^{\dag} (\widebar{z}_{3})  \widebar{V}_{[0,1]} (\widebar{z}_{2}) \widebar{V}_{[0,1]} (\widebar{z}_{1})  \widebar{\widetilde{s}}_{1}^{+}(\widebar{z})  \widetilde{s}_{1}^{+}(z) V_{[0,1]}(z_{1}) V_{[0,1]}(z_{2}) V_{[0,1]}(z_{3}) .
\end{equation}
The result at $z_{3} \to 0$ limit is 
\begin{equation*}
    \sum_{k} z_{12}^{\frac{1}{6}} z_{1}^{-\frac{1}{6}-k_{1}} z_{2}^{\frac{1}{2} +\sum k_{i}}    \widebar{z}_{12}^{\frac{1}{6}}  \widebar{z}_{1}^{k_{3}-\frac{5}{6}} \widebar{z}_{2}^{\frac{7}{6}+k_{4}} (1- \vert z_{1} \vert )^{\frac{1}{6}} (1- z_{1}\widebar{z}_{2})^{\frac{1}{6}}(1- z_{2}\widebar{z}_{1})^{\frac{1}{6}} (1- \vert z_{2} \vert )^{\frac{1}{6}} 
\end{equation*}
\begin{equation}
    \frac{\Gamma(2+k_{2})  \Gamma^{3}(\frac{2}{3})}{\Gamma(\frac{8}{3}+k_{2})\Gamma(\frac{4}{3})} \frac{(\frac{2}{3})_{\sum k_{i}} (\frac{1}{3})_{k_{1}} (\frac{4}{3})_{k_{2}} (\frac{1}{3})_{k_{3}} (\frac{1}{3})_{k_{4}} }{ k_{1}! k_{2}! k_{3}! k_{4}! (\frac{4}{3})_{\sum k_{i}}}  F_{D}^{(3)} \Big[ 2+k_{2}, \frac{1}{3}, \frac{1}{3}, \frac{1}{3} ; \frac{8}{3}+k_{2} ;\frac{\widebar{z}_{2}}{\widebar{z}_{1}} , z_{1} \widebar{z}_{2} , \vert z_{2}  \vert^{2} \Big]. 
\end{equation}
from which the power of the free-field approach is revealed when the number of the bulk operators is increased. The integral gets significantly more complicated, however, there is nothing technically difficult since the procedure is identical to those simple cases.

\

\paragraph{Level-$2$.}

There are a total of three dominant modules at level-$2$, hence the Kac-table contains $L_{[2,0]}$, $L_{[1,1]}$, and $L_{[0,2]}$. The modular $S$ matrix is 
\begin{equation}
    S= \frac{1}{2} \begin{pmatrix}
        1 & \sqrt{2} & 1 \\  \sqrt{2} & 0 & -\sqrt{2}  \\ 1  &  -\sqrt{2} & 1
    \end{pmatrix} ,
\end{equation}
with $S^{2}=C=I$. The fusion rules of the $\widehat{A}_{1}(2)$ theory are
\begin{equation}
    L_{[1,1]} \times L_{[1,1]} =L_{[2,0]} + L_{[0,2]} , \quad L_{[1,1]} \times L_{[0,2]} = L_{[1,1]} , \quad L_{[0,2]} \times L_{[0,2]} = L_{[2,0]}.
\end{equation}

\

We first consider the simple example $\langle b \vert \Phi_{[1,1]} (z_{1} , \widebar{z}_{1} ) \Phi_{[0,2]} (z_{2} , \widebar{z}_{2} ) \vert 0 \rangle $. For the Cardy boundary condition $ b_{[1,1]}$, the result is zero. For other Cardy boundary conditions, the disk correlation functions include the only contribution $I_{[1,1][0,2]}^{[1,1]}$. A free-field realization of it is 
\begin{equation}
    \oint dz\: \widebar{V}_{[0,2]}^{\dag} (\widebar{z}_{2}) \widebar{V}_{[1,1]}(\widebar{z}_{1}) \widetilde{s}^{+} (z) V_{[1,1]}(z_{1}) V_{[0,2]} (z_{2}). 
\end{equation}
The disk result is 
\begin{equation}
    \vert z_{1} \vert^{-1} (1-\vert z_{1} \vert^{2} )^{\frac{1}{8}} \frac{\Gamma(-\frac{1}{2})\Gamma(\frac{3}{4})}{\Gamma(\frac{1}{4})} \: _{1}F_{2} \Big[-\frac{1}{2} , \frac{1}{4} ; \frac{1}{4} ; \vert z_{1} \vert^{2} \Big] .
\end{equation}

\

A less simple example is the calculation of the $\langle b_{\widehat{\Lambda}} \vert \Phi_{[0,2]} (z_{1},\widebar{z}_{1}) \Phi_{[0,2]} (z_{2},\widebar{z}_{2}) \vert 0 \rangle$ and $\langle c \vert \Phi_{[0,2]} (z_{1},\widebar{z}_{1}) \Phi_{[0,2]} (z_{2},\widebar{z}_{2}) \vert 0 \rangle$, which consist of the contribution $I_{[0,2][0,2]}^{[2,0]}$. A free-field realization is 
\begin{equation}
     \oint dz^{(2)} \oint dz^{(1)} \: \widebar{V}_{[0,2]}^{\dag} (\widebar{z}_{2}) \widebar{V}_{[0,2]}(\widebar{z}_{1})\widetilde{s}^{+} (z^{(2)}) \widetilde{s}^{+} (z^{(1)}) V_{[0,2]}(z_{1}) V_{[0,2]} (z_{2}). 
\end{equation}
The disk result is 
\begin{equation}
   \sum_{k} \vert z_{1} \vert^{-2+2k_{2}} (1-\vert z_{1} \vert^{2})^{\frac{1}{2}} c( -\frac{1}{2} , \frac{1}{2} , \frac{1}{2} ;  1 ; k_{1},k_{2})  \frac{\Gamma(\sum_{i}k_{i}-\frac{1}{2}) \Gamma(\frac{1}{2})}{\Gamma(\sum_{i}k_{i})} \: _{1}F_{2} \Big[ \sum_{i}k_{i}-\frac{1}{2} , \frac{1}{2}   ; \sum_{i}k_{i}  ;  \vert z_{1} \vert^{2} \Big].
\end{equation}

\ 

\subsection{Disk correlations functions of $\widehat{A}_{2}(k)$ models}

The explicit forms of the two $\widehat{A}_{2}(k)$ screening operators are 
\begin{equation}
    \widetilde{s}_{1}^{+} = ( \beta^{1} -  \gamma^{2} \beta^{3} ) \: : e^{-i \alpha_{+} \alpha_{1} \phi } :\: , \quad  \widetilde{s}_{2}^{+} = ( \beta^{2} ) \: : e^{-i \alpha_{+} \alpha_{2} \phi } :\: .
\end{equation}

\

\paragraph{$\widehat{A}_{2}(1)$ models.}

The Kac-table includes three irreducible modules $\widehat{\Lambda}=[1,0,0]$, $[0,1,0]$, and $[0,0,1]$, with $UL_{[0,1,0]}=L_{[0,0,1]}$. The modular matrices are 
\begin{equation}
    S=  \frac{1}{\sqrt{3}} \begin{pmatrix}
        1 &  1  &   1 \\  1  & \kappa & \kappa^{2}  \\  1  & \kappa^{2} & \kappa 
    \end{pmatrix} ,\quad \kappa=e^{\frac{2\pi i} {3}} .
\end{equation}
The fusion rules are
\begin{equation}
    L_{[0,1,0]} \times L_{[0,0,1]} = L_{[1,0,0]}, \quad L_{[0,1,0]} \times L_{[0,1,0]} = L_{[0,0,1]} , \quad L_{[0,0,1]} \times L_{[0,0,1]} = L_{[0,1,0]} .
\end{equation}

\ 

Consider the correlation function $\langle b_{\widehat{\Lambda}} \vert \Phi_{[0,1,0]} (z_{1} , \widebar{z}_{1} ) \Phi_{[0,0,1]} (z_{2} , \widebar{z}_{2} ) \vert 0 \rangle$ in both the charge conjugated and the diagonal models.

For the charge conjugated model, a free field vertex operator realization of the contribution $^{U}I_{[0,1,0][0,0,1]}^{[1,0,0]}$ is 
\begin{equation}
   \oint dz^{(1)} \oint dz^{(2)} \widebar{V}_{[0,0,1]}^{\dag} (\widebar{z}_{2})  \widebar{V}_{[0,0,1]} (\widebar{z}_{1})  \widetilde{s}^{+}_{2} (z^{(2)}) \widetilde{s}^{+}_{1} (z^{(1)})  V_{[0,1,0]} (z_{1})   V_{[0,0,1]}  (z_{2}) .
\end{equation}
The result is 
\begin{equation}
     \sum_{k} \vert z_{1}\vert^{-\frac{4}{3}+2k} (1-\vert z_{1} \vert^{2})^{\frac{1}{12}} \Big[c(-\frac{1}{4} , \frac{1}{4} ; \frac{1}{2} ;k )  - c(\frac{3}{4} , \frac{1}{4} ; \frac{1}{2}  ; k)\Big] \frac{\Gamma(-\frac{1}{2}+k)\Gamma(\frac{3}{4})}{ \Gamma(\frac{1}{4}+k) } B(-\frac{1}{2} , \frac{3}{4}) .
\end{equation}

For the diagonal model, a free field vertex operator realization of the contribution $^{I}I_{[0,1,0][0,0,1]}^{[1,0,0]}$ is 
\begin{equation}
   \oint dz^{(1)} \oint dz^{(2)} \widebar{V}_{[0,1,0]}^{\dag} (\widebar{z}_{2})  \widebar{V}_{[0,1,0]} (\widebar{z}_{1})  \widetilde{s}^{+}_{1} (z^{(2)}) \widetilde{s}^{+}_{2} (z^{(1)})  V_{[0,1,0]} (z_{1})   V_{[0,0,1]}  (z_{2}) .
\end{equation}
The disk result is 
\begin{equation}
     \sum_{k} \vert z_{1}\vert^{-\frac{4}{3}+2k} (1-\vert z_{1} \vert^{2})^{\frac{1}{6}} \Big[c(-\frac{1}{4} , \frac{1}{4} ; \frac{1}{2} ;k )  - c(\frac{3}{4} , \frac{1}{4} ; \frac{1}{2}  ; k)\Big] \frac{\Gamma(-\frac{1}{2}+k)\Gamma(\frac{3}{4})}{ \Gamma(\frac{1}{4}+k) } B(-\frac{1}{2} , \frac{3}{4}) .
\end{equation}

\

\section{Discussion: A potential generalization to admissible level WZW models}

We close this work with a brief discussion on a potential generalization of the rational WZW models to the logarithmic WZW models, which is one of the most well-understood class of logarithmic CFT$_2$s \cite{Creutzig:2013hma}. Logarithmic CFT$_2$s are irrational CFT$_2$s, where logarithmic singularities appear in their correlation functions. It has been shown that the emerge of the logarithmic singularities is due to the involvement of the reducible yet indecomposable representations of the chiral algebras \cite{Knizhnik:1987xp, Saleur:1991hk, Saleur:1991vh, Rozansky:1992td, Gurarie:1993xq}.

I claimed that a CFT$_2$ admits a free-field approach, when free-field realizations of chiral algebra, Fock-space resolutions, and the calculation of correlation functions from free-field vertex operators can be given explicitly. The realization of the admissible-level Kac-Moody algebras is not difficult, where a direct reference is the free-field approaches to the rational quantum Drinfeld-Sokolov (QDS) $\mathcal{W}$ algebras. The main difference compared to the integer-level models is that there will be two background charged terms in the $\phi^{i}$ energy-stress tensor, together with a different set of screening operators $s^{-}_{i}  = \rho \: :e^{-i\alpha_{-} \alpha_{i} \cdot \phi }: (z)$, similar to the rational quantum Drinfeld-Sokolov (QDS) $\mathcal{W}$ minimal models \cite{Bouwknegt:1992wg}. The real challenge of generalizing to logarithmic CFT$_2$ begins with the reducible but indecomposable structures in their representation theories. In the past decades, there have been breakthroughs in the topic of realizing the relaxed module structure \cite{1711.11342, Kawasetsu:2018lur, Kawasetsu:2019att}. The construction of the free-field vertex operators in some of the admissible level WZW models have been given explicitly \cite{Creutzig:2013hma}, however, to make the method systematically applicable to all genus-zero correlation functions for the admissible level WZW models remains unsolved. Hence, this work, together with other few works on other types of RCFTs, can be considered as a medium step towards considering the same problems, in the modern developed logarithmic CFT$_2$s.

\

\acknowledgments

The author would like to thank H.Z.Liang for guidance and encouragement.

\end{document}